# Weather in stellar atmosphere: The dynamics of mercury clouds in $\alpha$ Andromedae

Oleg Kochukhov[1], Saul J. Adelman[2,4], Austin F. Gulliver[3,4] & Nikolai Piskunov[1]

[1]*Department of Astronomy and Space Physics, Uppsala University, Box 515, SE 75120 Uppsala, Sweden*
[2]*Department of Physics, The Citadel, 171 Moultrie Street, Charleston, SC 29409, USA*
[3]*Department of Physics and Astronomy, Brandon University, Brandon, MB, R7A 6A9, Canada*
[4]*Guest Investigator, Dominion Astrophysical Observatory, Herzberg Institute of Astrophysics, National Research Council of Canada, Canada*

**The formation of long-lasting structures at the surfaces of stars is commonly ascribed to the action of strong magnetic fields. This paradigm is supported by observations of evolving cool spots in the Sun[1] and active late-type stars[2], and stationary chemical spots in the early-type magnetic stars[3]. However, results of our seven-year monitoring of mercury spots in non-magnetic early-type star $\alpha$ Andromedae show that the picture of magnetically-driven structure formation is fundamentally incomplete. Using an indirect stellar surface mapping technique, we construct a series of 2-D images of starspots and discover a secular evolution of the mercury cloud cover in this star. This remarkable structure formation process, observed for the first time in any star, is plausibly attributed to a non-equilibrium, dynamical evolution of the heavy-element clouds created by atomic diffusion[4] and may have the same underlying physics as the weather patterns on terrestrial and giant planets.**

Since the detection of sunspot magnetic fields[5] it has become clear that magnetic forces are responsible for the activity and the formation of surface structure in the Sun. Strong magnetic fields, which are generated in the solar interior and brought to the surface by convective motions, inhibit the outward flow of energy, thus creating prominent dark spots with filamentary structure[1]. A very similar relation between magnetic fields and cool surface spots is observed in many active solar-type stars[2]. All these objects possess highly structured magnetic fields, which evolve on time scales from weeks to months. Stars significantly more massive than the Sun do not have an energetic surface convection zone. The lack of mixing in their atmospheres allows diffusive segregation of chemical elements to operate efficiently. The action of the competing forces of downward gravitation acceleration and the outward radiation pressure, collectively referred to as atomic diffusion[6], leads to selective heavy-element enrichment of their stellar atmospheres. This phenomenon of chemical peculiarity is present in most slowly rotating A and B-type stars. The presence of strong, well-organized magnetic fields in some chemically peculiar (CP) stars alters the efficiency of the atomic transport processes, resulting in the formation of non-uniform distributions of chemical abundances over the surface and with height in their atmospheres. The fields and chemical spots in magnetic CP stars are essentially static, showing no evolutionary changes on the time scales accessible to human observers.

Despite a qualitative difference between the magnetic characteristics and related variability patterns of the low-mass solar-type stars and massive early-type objects, an undeniable and seemingly universal link between the presence of magnetic fields and the surface structure formation exists in both groups of stars. A series of recent studies has challenged this picture. A small group of late-B CP stars of mercury-manganese (HgMn) peculiarity type has no convincing evidence of the dynamically important magnetic fields[7] but, at the same time, exhibits clear signatures of a non-uniform surface distribution of mercury and several other heavy elements[8–10]. The bright HgMn star $\alpha$ And (HR 15, HD 358) was the first non-magnetic spotted star discovered. An extensive magnetic field search[11] has established at the 99% confidence level that no fields stronger than 50–100 G are present in the atmosphere of this star. This upper limit is significantly smaller than the field strength for which the magnetic and thermodynamic energy densities are equal (equipartition field, $\approx 250$ G for the lower atmospheric layers). This finding effectively rules out any magnetically-driven creation of chemical spots. Thereby $\alpha$ And and other spotted HgMn stars present a formidable theoretical and observational challenge to our understanding of the formation of structure in stellar atmospheres.

What mechanism may be responsible for the origin of spots in non-magnetic stars can be clarified with a comprehensive investigation, which examines the two-dimensional geometry of the surface chemical inhomogeneities and probes their possible temporal evolution. We undertook such a study for the Hg spots in $\alpha$ And. This star was observed in a seven-year-long campaign at the Dominion Astrophysical Observatory in Canada and the Special Astrophysical Observatory in Russia. Spectroscopic monitoring focused on the strongest Hg II line (398.4 nm) in the optical wavelength region of $\alpha$ And. This transition exhibits conspicuous variation in line strength and line profile shape, caused by the rotational modulation of the geometrical aspect at which we observe a non-axisymmetric surface distribution of this chemical element.

We use the equivalent width measurements of the Hg II 398.4 nm line to refine the rotation period of $\alpha$ And and to quantify the short- and long-term Hg II line strength behaviour. The line strength data folded with rotation phase are illustrated in Fig. 1a. The peak-to-peak amplitude of the equivalent width modulation reaches 40%. The overall shape of the line strength curve is stable over the period of seven years, but a notable change from one epoch to another is detected close to the rotation phase 0.5, corresponding to the maximum in equivalent width. Fourier analysis (Fig. 1b) of the Hg II equivalent width curves constructed for the three epochs with complete coverage of the rotation cycle shows that the discrepancy in the line strength behaviour observed for different epochs is highly significant ($> 99$% confidence level). A complimentary investigation of the Hg II line profile morphology (see Supplementary Fig. 1) confirms that the observed modification of the equivalent width curve corresponds to a real change in the line profile variability pattern. We attribute this secular variation of the Hg II 398.4 nm line strength curve to a long-term evolution of the mercury spot structure in $\alpha$ And.

To obtain a complete picture of the Hg spot dynamics in $\alpha$ And, we applied a Doppler imaging procedure[8] to the subsets of the Hg II 398.4 nm spectrograms collected in 1998, 2002 and 2004. This enabled us to construct a series of 2-D maps of the stellar surface, giving insight into the evolution of the Hg inhomogeneities. The three independently reconstructed mercury maps are presented in Fig. 2a. The common feature of these images is a significant pole-to-equator gradient in chemical abundance. The mercury concentration at the rotation poles of $\alpha$ And exceeds the solar abundance of this element by a factor of 250, whereas the typical overabundance in the equatorial region is $2.5 \times 10^6$. Several large (30°–60°) mercury clouds at low and intermediate latitudes distort the axisymmetric pattern, leading to the observed rotational modulation of the Hg II line. The structure of the mercury clouds clearly changes on a time scale of 2–4 years (Fig. 2b). In particular, a comparison of the 1998-year map with the abundance distributions derived for years 2002 and 2004 shows the emergence of a new large Hg spot at longitude $\approx 250°$. At the same time, the concentration of mercury has decreased around longitude 330°. The amplitude of these abundance variations reaches a factor of 100 in some surface regions, which is comparable to the star-to-star scatter in the mean level of the Hg abundance in the atmospheres of HgMn stars[12].

The discovery of the secular variation of the Hg spot geometry in $\alpha$ And opens an entirely new perspective on structure formation mechanisms in stellar atmospheres. Our analysis is the first to reveal evolving chemical inhomogeneities, apparently unrelated to magnetic fields, in any star. Radiative diffusion calculations[4] predict that Hg accumulates in a thin cloud in the upper atmosphere, where the gravitational force and radiative pressure nearly cancel out. Levitating Hg atoms become susceptible to minute external perturbations. A 1.3% decrease in the effective gravity due to the centrifugal force at the equator of $\alpha$ And corresponds to a roughly 300 times higher Hg abundance. A small upward displacement of the mercury cloud converts Hg II atoms to the unobservable Hg III state, thus substantially reducing the strength of the Hg II 398.4 nm line. In this scenario the observed time-dependent non-axisymmetric component of the chemical images is attributed to hydrodynamical instabilities which develop in the Hg-rich layer. The Hg cloud cover could be disrupted by the thermohaline instability induced by inverse molecular weight gradients[13] or by the dynamic tidal perturbation[14] due to the secondary star moving in a highly eccentric orbit. Alternatively, a time-dependent radiative diffusion scenario[15], involving several interacting Hg layers at different heights, could contribute to the variation of the chemical structures in $\alpha$ And. Independently of the theory scrutinizing details of the atmospheric diffusion instabilities, one can draw a parallel between evolving chemical cloud "weather" in stars and the weather patterns on the Earth and the giant gas planets. Our investigation demonstrates that, similar to planetary atmospheres, the outermost layers of stars are sites of dynamical processes of structure formation and self-organization.

The observation of the Hg cloud dynamics in $\alpha$ And offers a possible solution to the long-standing problem of the chemical diversity of HgMn and related stars. If a competition between radiative levitation and gravity, with the possible influence of weak mass loss[16], is the primary mechanism responsible for the accumulation of heavy elements in the atmospheres of HgMn and similar CP stars, one would expect to find very similar abundances in stars of the same mass and evolutionary stage. Contrary to this prediction of the simple diffusion theory, a large discrepancy in the heavy[12] and iron-peak[17] element abundances is often reported for stars with similar funda-mental parameters. We propose that the heavy-element atmospheric enrichment created by atomic diffusion is not stationary but evolves under the influence of the hydrodynamical instability of the same kind as observed in $\alpha$ And. As a result of this time-variable process, atmospheric abundances change quickly compared to the stellar evolutionary time scale. Thus, the observed concentrations of the elements most sensitive to the diffusion effect include a significant stochastic or, perhaps, a cyclic time-dependent component and are only loosely related to the stellar parameters. The mere presence and the magnitude of the characteristic chemical peculiarities may represent a transient phase in the long-term evolution of the chemical composition of the stellar surface layers.

**Methods**

**Spectroscopic observations.** We have collected 127 observations of $\alpha$ And in the Hg II 398.4 nm region during five different epochs, spanning from 10 days to 4 months, over the period of seven years. Observations were obtained with the high-resolution spectrographs attached to the 1.2-m telescope of the Dominion Astrophysical Observatory (Canada) and the 6-m telescope of the Special Astrophysical Observatory (Russia). The spectra have resolving powers of $\lambda/\Delta\lambda = 36{,}500$–$44{,}000$ and signal-to-noise (S/N) ratios between 500:1 to 1500:1. Reduction of the spectrograms utilized our optimal extraction programmes[18,19] to optimize the S/N of the $\alpha$ And observations.

**Time-series analysis.** Confirming results of the previous studies[8], we find that the Hg II 398.4 nm line exhibits remarkable variability in the average intensity and profile shape on a time scale of $\approx 2.38$ d. Lines of other chemical elements show no evidence of the modulation caused by surface spots. The stellar rotation period of $2.38195 \pm 0.00003$ d is inferred from the changes in Hg II equivalent width. Two independent time series analysis methods, a third-order least-squares Fourier fitting and Phase Dispersion Minimisation[20], yield consistent values of the rotation period.

**Stellar parameters.** $\alpha$ And is a spectroscopic binary star with a spotted HgMn B8p primary and an A3m secondary[21,22]. The orbital period is 96.70 d. The atmospheric parameters[22] essential for modelling the spectra of the system are $T_{\rm eff} = 13{,}800$ K, $\log g = 3.75$ for the primary and $T_{\rm eff} = 8{,}500$ K, $\log g = 4.0$ for the secondary, respectively. The less massive star is 2 mag fainter than the primary and has a much higher projected rotational velocity. Contribution from its line absorption is negligible at the location of the Hg II 398.4 nm feature. However, the secondary's contribution to the continuum radiation weakens the lines of the primary star. We compensate for this effect by correcting the observed spectra: $S_{\rm cor} = S(1 + 1/R) - 1/R$, where the luminosity ratio $R \equiv L_{\rm prim}/L_{\rm sec} = 8.72$ at 398.4 nm.

**Doppler imaging.** We reconstruct the distribution of the Hg abundance over the surface of $\alpha$ And with the help of Doppler imaging[23]. This technique takes advantage of the partial resolution of the stellar surface provided by the rotational Doppler effect, inverting a line profile time series into a 2-D map of the stellar surface. The abundance Doppler imaging code INVERS12[3] is applied to interpret the observations of $\alpha$ And. The stellar line profiles are computed on a $30 \times 60$-element surface grid, using ATLAS9[24] model atmosphere. The projected rotational ve-locity, $v \sin i = 53.0 \pm 0.5$ km s$^{-1}$, is determined from the line profile fit, and the inclination angle $i = 75^\circ$ is adopted following refs. [4,21]. The line list used for spectrum synthesis calculations includes 15 isotope and hyperfine components[12] of the Hg II 398.4 nm transition and the Y II 398.26 nm line, which blends with the blue wing of the Hg II feature. The hydrogen bound-bound opacity due to the H$\varepsilon$ line at 397 nm is also accounted for. In our surface mapping procedure the Hg isotope mixture is adjusted simultaneously with the total abundances of these elements. We find that the isotope composition of Hg in $\alpha$ And does not deviate significantly from the solar mixture and varies little over the surface of the star. A comparison of the observed and computed Hg II line profiles (Supplementary Figure 2) indicates that our models of the Hg surface spots yield a good fit to observations for all three epochs considered.

**Acknowledgements**   O.K. and N.P. acknowledge financial support from the Swedish Research Council and the Royal Academy of Sciences. Drs. M. Sachkov and D. Kudryavtsev are thanked for their assistance with observations of $\alpha$ And at the 6-m SAO telescope. S.A. and A.G. thank Dr. James E. Hesser, Director of the Dominion Astrophysical Observatory for the observing time. S.A.'s contribution to this paper was supported in part by grants from The Citadel Foundation. Financial support was provided to A.G. by the Natural Sciences and Engineering Research Council of Canada.

**Competing Interests**   The authors declare that they have no competing financial interests.

**Correspondence**   Correspondence and requests for materials should be addressed to O.K. (email: Oleg.Kochukhov@astro.uu.se).

**Figure 1** The Hg II 398.4 nm line strength variation over the rotation cycle of $\alpha$ And. **a**, equivalent width measurements for five different epochs are plotted as a function of rotation phase. **b**, a third-order Fourier fit to the line strength variation in years 1998, 2002 and 2004. The width of the curves corresponds to a $2\sigma$ (95% confidence level) uncertainty of the approximating function. The significant discrepancy of the curves in the 0.3–0.7 rotational phase interval demonstrates that the surface distribution of mercury, responsible for the Hg II line variability, is undergoing a slow evolution.

**Figure 2** Distribution of the Hg clouds over the surface of $\alpha$ And in 1998, 2002 and 2004. **a**, the Hammer-Aitoff projection of the 2-D Hg maps reconstructed with Doppler imaging. The dotted curves are lines of constant longitude and latitude, plotted every $30°$. The longitude increases from left to right, covering the range of $0°$–$360°$. Darker regions represent higher element abundance. The Hg concentration is given relative to the solar value: $[\text{Hg}] = \log(N_{\text{Hg}}/N_{\text{H}}) - \log(N_{\text{Hg}}/N_{\text{H}})_\odot$, where $\log(N_{\text{Hg}}/N_{\text{H}})_\odot = -10.87$[25]. **b**, the Hg difference maps showing the variation of the mercury surface distribution from one epoch to another.

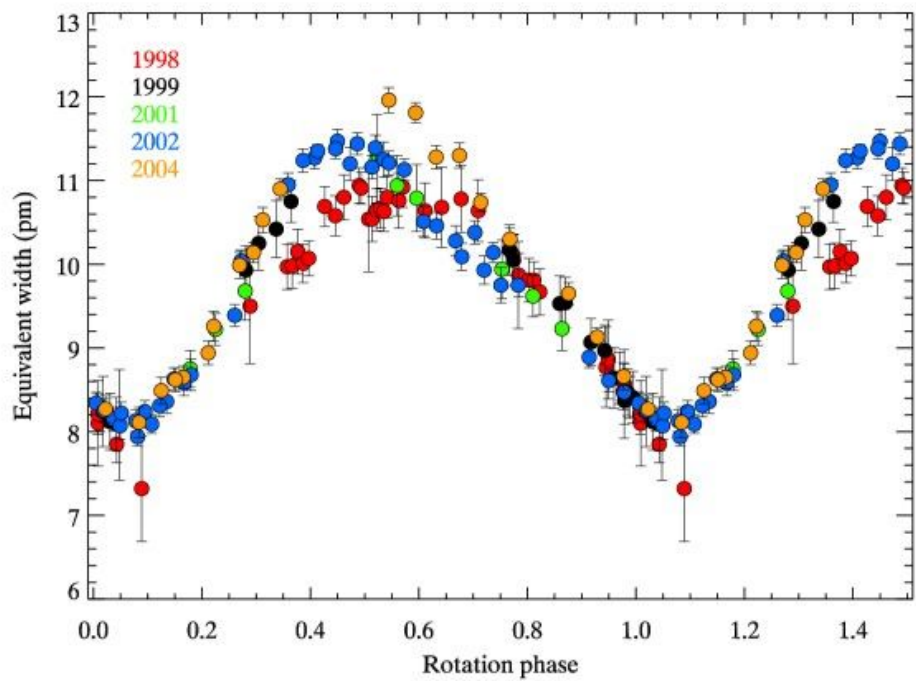

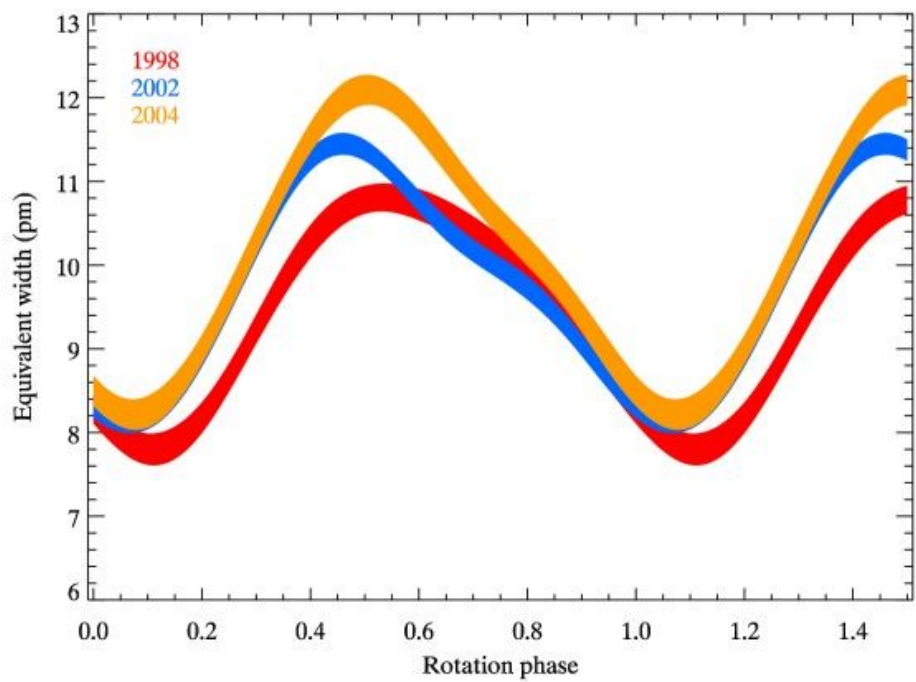

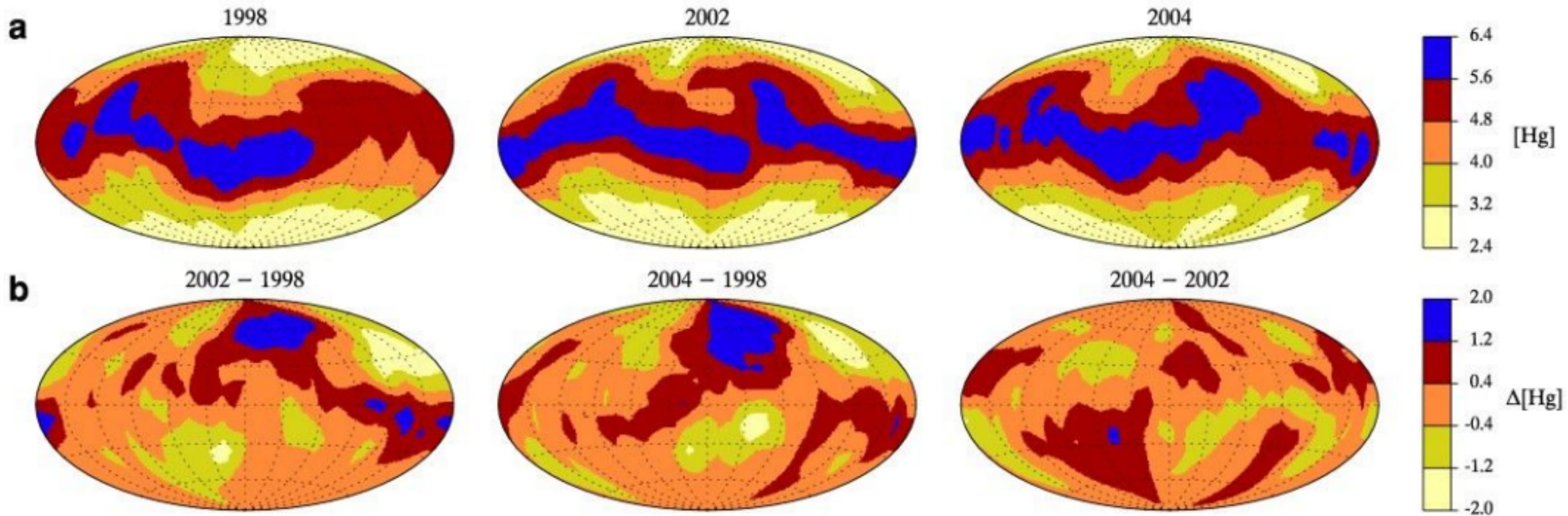